\documentclass{amsart}
\usepackage[dvips]{graphicx}
\usepackage{hyperref}
\graphicspath{ {.} }
\setkeys{Gin}{width=\linewidth,keepaspectratio=true}
\usepackage[latin1]{inputenc}
\usepackage{natbib}
\bibpunct{(}{)}{;}{a}{,}{;}
\usepackage{color}
\usepackage{url}
\usepackage{epsfig}
\usepackage{algorithm}
\usepackage{algpseudocode}
\usepackage{rotating}
\usepackage{lscape}
\usepackage{amsmath,amssymb}
\usepackage{numprint}
\npdecimalsign{.} 

\usepackage{mfirstuc}

\newcommand{\Xcomment}[1]{}

\begin{document}

\title[New Instances for Maximum Weight Independent Set]{New Instances for Maximum Weight Independent\\ Set
  From a Vehicle Routing Application}

\author[Y. Dong]{Yuanyuan Dong}

\address[Yuanyuan Dong]{Dallas, TX\ USA. The work was done while the author was at Amazon.com}

\email[]{njdyy03@gmail.com}

\author[A.V. Goldberg]{Andrew V. Goldberg}

\address[Andrew V. Goldberg]{Amazon.com,
East Palo Alto, CA\ USA.}

\email[]{\href{mailto:avg@alum.mit.edu}{avg@alum.mit.edu}}

\author[A. Noe]{Alexander Noe}

\address[Alexander Noe]{University of Vienna, Vienna, Austria.
  The work was done while the author was at Amazon.com}

\email[]{alexander.noe@univie.ac.at}

\author[N. Parotsidis]{Nikos Parotsidis}

\address[Nikos Parotsidis]{
Google, Zurich, Switzerland. The work was done while the author was at Amazon.com}

\email[]{nickparo1@gmail.com}

\author[M.G.C. Resende]{Mauricio G.C. Resende}

\address[Mauricio G.C. Resende]{Amazon.com and Industrial \& 
Systems Engineering,
University of Washington, Seattle, WA\ USA.}
\email[]{mgcr@berkeley.edu, mgcr@uw.edu}

\author[Q. Spaen]{Quico Spaen}

\address[Quico Spaen]{Amazon.com,
East Palo Alto, CA\ USA.}

\email[]{qspaen@berkeley.edu}

\begin{abstract}
We present a set of new instances of the maximum weight independent set problem.
These instances are derived from a real-world vehicle routing problem and are 
challenging to solve in part because of their large size.
We present instances with up to 881 thousand nodes and 383 million edges.
\end{abstract}

\date{\today --
Parts of this paper were written while the
first, third, and fourth authors were employed at
Amazon.com.}

\maketitle

\section{Vehicle Routing Application of MWIS}
Given an undirected graph $G=(V,E)$ where $V$ is its set of nodes and $E$ its 
set of edges, a subset of nodes $S \subseteq V$ is an \textit{independent set}
if the elements of $S$ are pairwise nonadjacent in $G$.
If $w(v)$ is the weight of node $v \in V$, the weight of independent set $S$ is
$W(S) = \sum_{v \in S} w(v)$.
In the \textit{maximum weight independent set} (MWIS) problem we seek an independent
set $S^*$ such that $W(S^*) \geq W(S)$ for all independent sets $S \subseteq V$
in $G$.
This optimization problem is NP-hard \citep{GarJoh79a} and it is often solved 
using heuristic algorithms.

We provide a collection of instances of an
MWIS problem that appeared as subproblems in algorithms solving
real-life long-haul vehicle routing problems at Amazon.
Our goal is to enhance the set of benchmark instances available to
algorithm researchers working on MWIS.
Our instances differ from other publicly available
instances and the new collection includes some large instances.

To gain intuition into the application, consider a stochastic heuristic for
the problem.
This heuristic produces different solutions for different
pseudo-random generator seeds.
Each solution consists of a set of routes.
We want to recombine routes from multiple solutions to obtain a better
solution.

Each {\em route} consists of a driver and a set of loads assigned to the driver.
A subset of routes is {\em feasible} if no two routes in the subset share
a driver or a load.
Each route has a weight.
The objective function is the sum of route weights.
The problem is to find a feasible solution of the maximum total weight.

To state this problem as MWIS, we build a {\em conflict graph} as follows.
Nodes of the graph correspond to routes and weights correspond to route
weights.
We connect two nodes by an edge if the corresponding routes have a 
conflict, i.e.,
they share a driver or a load.

\begin{table}
    \caption{\label{t:gr_vr} List of VR instances in the library.  For each 
of the 38 instances, the table lists the instance name, the number of nodes and
edges in the conflict graph, the total weight of a starting solution, the 
linear programming (LP) upper bound, the compressed tar files of the directory
with the files that define the instance, and the size (in Mbytes) of the 
compressed tar file.}
{\tiny
    \begin{tabular}{l | rr | r | r | r | r}
        Instance & $|V|$ & $|E|$ & Initial Sol. & LP bound & Filename & Mbytes \\\hline
&&&&&&\\
        MT-D-01 &
 \numprint{979} & \numprint{3841} & \numprint{228874404} & \numprint{238166485} &
        \texttt{MT-D-01.tar.gz} & \numprint{0.03} \\
        MT-D-200 &
 \numprint{10880} & \numprint{547529} & \numprint{286750411} & \numprint{287228467} &
        \texttt{MT-D-200.tar.gz} & \numprint{1.77} \\
        MT-D-FN &
 \numprint{10880} & \numprint{645026} & \numprint{290723959} & \numprint{290881566} &
        \texttt{MT-D-FN.tar.gz} & \numprint{2.07} \\
        MT-W-01 &
 \numprint{1006} & \numprint{3140} & \numprint{299132358} & \numprint{312121568} &
        \texttt{MT-W-01.tar.gz} & \numprint{0.03}
\\
        MT-W-200 &
 \numprint{12320} & \numprint{554288} & \numprint{383620215} & \numprint{384099118} &
        \texttt{MT-W-200.tar.gz} & \numprint{1.86} \\
        MT-W-FN &
 \numprint{12320} & \numprint{593328} & \numprint{390596383} & \numprint{390869891} &
        \texttt{MT-W-FN.tar.gz} & \numprint{1.97} \\  
&&&&&&\\
        MR-D-01 &
 \numprint{14058} & \numprint{60738} & \numprint{1664446852} & \numprint{1695332636} & 
        \texttt{MR-D-01.tar.gz} &
\numprint{0.48}
\\
        MR-D-03 &
 \numprint{21499} & \numprint{168504} & \numprint{1739544141} & \numprint{1763685757} &
        \texttt{MR-D-03.tar.gz} &
\numprint{0.97}
\\
        MR-D-05 &
 \numprint{27621} & \numprint{295700} & \numprint{1775123794} & \numprint{1796703313} &
        \texttt{MR-D-05.tar.gz} &
\numprint{1.35}
\\
        MR-D-FN &
 \numprint{30467} & \numprint{367408} & \numprint{1794070793} & \numprint{1809854459} &
        \texttt{MR-D-FN.tar.gz} &
\numprint{1.75}
\\
        MR-W-FN &
 \numprint{15639} & \numprint{267908} & \numprint{5386472651} & \numprint{5386842781} &
        \texttt{MR-W-FN.tar.gz} &
\numprint{1.18}
\\ 
&&&&&&\\
        MW-D-01 &
 \numprint{3988} & \numprint{19522} & \numprint{465730126} & \numprint{477563775} &
        \texttt{MW-D-01.tar.gz} & 
\numprint{0.14}
\\
        MW-D-20 &
 \numprint{20054} & \numprint{718152} & \numprint{522485254} & \numprint{531510712} &
        \texttt{MW-D-20.tar.gz} &
\numprint{2.50}
\\
        MW-D-40 &
 \numprint{33563} & \numprint{2169909} & \numprint{533938531} & \numprint{543396252}  &
        \texttt{MW-D-40.tar.gz} &
\numprint{7.20}
\\
        MW-D-FN &
 \numprint{47504} & \numprint{4577834} & \numprint{542182073} & \numprint{549872520}  &
        \texttt{MW-D-FN.tar.gz} &
\numprint{15.17}
\\
        MW-W-01 &
 \numprint{3079} & \numprint{48386} & \numprint{1268370807} & \numprint{1270311626} &
        \texttt{MW-W-01.tar.gz} &
\numprint{0.21}
\\
        MW-W-05 &
 \numprint{10790} & \numprint{789733} & \numprint{1328552109} & \numprint{1334413294} &
        \texttt{MW-W-05.tar.gz} &
\numprint{2.49}
\\
        MW-W-10 &
 \numprint{18023} & \numprint{2257068} & \numprint{1342415152} & \numprint{1360791627} &
        \texttt{MW-W-10.tar.gz} &
\numprint{6.76}
\\
        MW-W-FN &
 \numprint{22316} & \numprint{3495108} & \numprint{1350675180} & \numprint{1373020454} & 
        \texttt{MW-W-FN.tar.gz} &
\numprint{10.41}
\\ 
&&&&&&\\
        CW-T-C-1 &
 \numprint{266403} & \numprint{162263516} & \numprint{1298968} & \numprint{1353493} &
        \texttt{CW-T-C-1.tar.gz} &
\numprint{547.73}
\\
        CW-T-C-2 &
 \numprint{194413} & \numprint{125379039} & \numprint{933792} & \numprint{957291} &
        \texttt{CW-T-C-2.tar.gz} &
\numprint{417.49}
\\
        CW-T-D-4 &
 \numprint{83091} & \numprint{43680759} & \numprint{457715} & \numprint{463672} &
        \texttt{CW-T-D-4.tar.gz} &
\numprint{140.88}
\\
        CW-T-D-6 &
 \numprint{83758} & \numprint{44702150} & \numprint{457605} & \numprint{463946} &
        \texttt{CW-T-D-6.tar.gz} &
\numprint{143.95}
\\ 
&&&&&&\\
        CR-T-C-1 &
 \numprint{602472} & \numprint{216862225} & \numprint{4605156} & \numprint{4801515} & 
        \texttt{CR-T-C-1.tar.gz} &
\numprint{746.32}
\\
        CR-T-C-2 &
 \numprint{652497} & \numprint{240045639} & \numprint{4844852} & \numprint{5032895} &
        \texttt{CR-T-C-2.tar.gz} &
\numprint{828.21}
\\
        CR-T-D-4 &
 \numprint{651861} & \numprint{245316530} & \numprint{4789561} & \numprint{4977981} &
        \texttt{CR-T-D-4.tar.gz} &
\numprint{845.85}
\\
        CR-T-D-6 &
 \numprint{381380} & \numprint{128658070} & \numprint{2953177} & \numprint{3056284} &
        \texttt{CR-T-D-6.tar.gz} &
\numprint{441.42}
\\
        CR-T-D-7 &
 \numprint{163809} & \numprint{49945719} & \numprint{1451562} & \numprint{1469259} &
        \texttt{CR-T-D-7.tar.gz} &
\numprint{168.95}
\\ 
&&&&&&\\
        CW-S-L-1 &
 \numprint{411950} & \numprint{316124758} & \numprint{1622723} & \numprint{1677563} &
        \texttt{CW-S-L-1.tar.gz} &
\numprint{1071.34}
\\
        CW-S-L-2 &
 \numprint{443404} & \numprint{350841894} & \numprint{1692255} & \numprint{1759158} &
        \texttt{CW-S-L-2.tar.gz} &
\numprint{1192.32}
\\
        CW-S-L-4 &
 \numprint{430379} & \numprint{340297828} & \numprint{1709043} & \numprint{1778589} &
        \texttt{CW-S-L-4.tar.gz} & \numprint{1156.28}
\\
        CW-S-L-6 &
 \numprint{267698} & \numprint{191469063} & \numprint{1159946} & \numprint{1192899} &
        \texttt{CW-S-L-6.tar.gz} &
\numprint{644.49}
\\
        CW-S-L-7 &
 \numprint{127871} & \numprint{89873520} & \numprint{589723} & \numprint{599271} &
        \texttt{CW-S-L-7.tar.gz} &
\numprint{294.53}
\\ 
&&&&&&\\
        CR-S-L-1 &
 \numprint{863368} & \numprint{368431905} & \numprint{5548904} & \numprint{5768579} & 
        \texttt{CR-S-L-1.tar.gz} &
\numprint{1271.78}
\\
        CR-S-L-2 &
 \numprint{880974} & \numprint{380666488} & \numprint{5617351} & \numprint{5867579} &
        \texttt{CR-S-L-2.tar.gz} &
\numprint{1314.11}
\\
        CR-S-L-4 &
 \numprint{881910} & \numprint{383405545} & \numprint{5629351} & \numprint{5869439} &
        \texttt{CR-S-L-4.tar.gz} &
\numprint{1323.34}
\\
        CR-S-L-6 &
 \numprint{578244} & \numprint{245739404} & \numprint{3841538} & \numprint{3990563} &
        \texttt{CR-S-L-6.tar.gz} &
\numprint{845.81}
\\
        CR-S-L-7 &
 \numprint{270067} & \numprint{108503583} & \numprint{1969254} & \numprint{2041822} &
        \texttt{CR-S-L-7.tar.gz} &
\numprint{370.47}
\\
&&&&&&\\
        \hline
    \end{tabular}
}

\end{table}
Our application has additional information that one can (optionally) use
in an algorithm.
First, we have a good initial solution,
the best of the solutions we combine.
We provide initial solutions for our instances.
One can use this solution to possibly warm-start a MWIS algorithm.

Second, we have information about many cliques in the conflict graph.
For a fixed load (or driver), nodes corresponding to the routes
containing the load (driver) form a clique: every pair of such nodes is
connected.
This allows us to use the well-known clique integer linear programming (ILP)
formulation of the problem:
\begin{align}
\max & \sum_{v\in V} w_v x_v \notag \\
\text{subject to} & \notag \\
& C_2, C_3, \ldots, C_k, \notag \\
& x_v \in \{0,1\}, \forall \; v \in V, \notag
\end{align}
where 
$C_2, C_3, \ldots, C_k$ are, respectively, the sets of 2-clique, 
3-clique$, \ldots,$ and  $k$-clique inequalities.
In general, for cliques $Q$ of size $k$, we have the set of $k$-clique 
inequalities 
\begin{displaymath}
\sum_{ v \in Q} x_v \leq 1,\; \mbox{for all cliques $Q$ of size $k$}.
\end{displaymath}

One can solve a linear programming (LP) relaxation of the problem, 
which assigns each node
a value in the closed real interval $[0,1]$.
Note that the objective function of the LP relaxation provides an upper bound
on the corresponding MWIS solution value.
We provide both the cliques and the relaxed LP solutions with our instances.

Table~\ref{t:gr_vr} lists the instances we provide and includes
the graph size, the initial solution value, and the relaxed LP bound.

\section{Input Graph Format}
We place each instance in a separate directory containing several files with
instance name, graph edge set, node weights, clique information, and relaxed LP
solution values.
Directory names correspond to the instance names.
Next we describe the file formats.

For an undirected, node-weighted graph $G=(V,E,w)$ with $n$ nodes, $m$ edges
and integral node IDs from $[1, n]$, we give the following files:

\begin{itemize}
\item \texttt{instance\_name.txt} -- Name of the instance.
\item \texttt{conflict\_graph.txt} -- Edges of $G$.
  The file has a total of $m + 1$ lines.
  The first line gives the numbers of nodes and edges: ``$n\;m$".
  Each of the lines $2, \ldots, m+1$ describes an edge $e = (u,v) \in E$
  as ``$u\;v$".
\item \texttt{node\_weights.txt} -- Node weights.
  The file has a total of $n$ lines, each describing the weight of node
  $v \in V$ as ``$v\;w(v)$''.
  The weights are integers.
\item \texttt{solution.txt} -- Initial solution for warm start.
  It contains one line per node in the initial solution, giving its node index:
  if a node $v$ in the solution, the file contains a line with ``$v$'' in it.
\item \texttt{cliques.txt} -- Set of cliques in $G$. 
For each clique $C=\{c_1,c_2,\dots,c_k\}$,
  the file contains one line as ``$c_1\;c_2\;\dots\;c_k$''.
\item \texttt{lploads.txt} -- Solution for the relaxed LP
  problem for the MWIS problem on the clique graph, where each node 
$v \in V$ has a relaxed LP value $l(v) \in [0,1]$. 
The file has $n$ lines, each with the LP value of a node $v\in V$ 
as ``$v\;l(v)$", where $l(v)$ is a floating point number.
\end{itemize}

The files
\texttt{conflict\_graph.txt} and \texttt{node\_weights.txt}
are needed by any MWIS algorithm.
The other files are optional.

Note that some of our graphs are large, with the compressed tar file being
over 1 Gbyte in size. 32-bit integers are insufficient
to represent the total weight of a solution.
An implementation needs to use 64-bit integers or doubles to represent the
weight of these independent sets.


\section{Downloading the Instances}

The full set of 38 instances can be downloaded 
as gzipped tar files
from the AWS OpenData site:
\vspace{15pt}

\noindent
\hspace{20pt} \url{https://registry.opendata.aws/mwis-vr-instances/}

\vspace{10pt}
\noindent
using the AWS command line interface (CLI) \citep{AWSCLI}.

Instruction on installation of AWS CLI can be found in \citet{AWSCLI}.  
As an example, installation on MacOS can be done using Terminal with
the commands

\vspace{10pt}
\begin{verbatim}
curl "https://awscli.amazonaws.com/AWSCLIV2.pkg" -o "AWSCLIV2.pkg"
sudo installer -pkg AWSCLIV2.pkg -target /
\end{verbatim}

\vspace{10pt}
To list the contents of the repository using AWS CLI, run the command

\vspace{10pt}
\begin{verbatim}
aws s3 ls s3://mwis-vr-instances/ --no-sign-request
\end{verbatim}

\vspace{10pt}
To download an instance, say file \texttt{MT-W-01.tar.gz}, from the 
repository using AWS CLI, run the command

\vspace{10pt}
\begin{verbatim}
aws s3 cp s3://mwis-vr-instances/MT-W-01.tar.gz . --no-sign-request
\end{verbatim}
\section{Concluding Remarks}

In this paper we introduce a set of large-scale maximum weight independent 
set instances arising in a real-world vehicle routing application. 
Our goal in making these instances available to other researchers is that
progress can be made in the field.
Other researchers can try their existing MWIS solvers on these instances 
and can be motivated to
develop new solvers for them. 

\section*{Acknowledgment}

On behalf of all authors, the corresponding author (MGCR) states that there 
is no conflict of interest, no funding, and the data is freely available..

\bibliographystyle{plainnat}
\bibliography{bibliography}   

\end{document}